\documentstyle[12pt,epsf]{article}
\def \bea{\begin{eqnarray}}

\def \beq{\begin{equation}}

\def \cst{\cos^2 \theta}

\def \eea{\end{eqnarray}}
\def \eeq{\end{equation}}

\def \g{{\rm~GeV}}

\def \ite{{\it et al.}}

\def \pr{\parallel}

\def \s{\sqrt{2}}
\def \sef{\sin^2 \theta^{\rm eff}}

\def \sst{\sin^2 \theta}

\def \SUL{SU(2)$_L$}

\def \U1Y{U(1)$_Y$}
\def \vev#1{\langle #1 \rangle}

\begin{document}
\centerline {\bf ROLE OF PRESENT AND FUTURE ATOMIC PARITY VIOLATION}
\centerline{\bf EXPERIMENTS IN PRECISION ELECTROWEAK TESTS
\footnote{Enrico Fermi Institute preprint EFI 01-43, hep-ph/0109239.
To be submitted to Physical Review D.}}
\bigskip

\centerline{Jonathan L. Rosner~\footnote{rosner@hep.uchicago.edu}}
\centerline {\it Enrico Fermi Institute and Department of Physics}
\centerline{\it University of Chicago, 5640 S. Ellis Avenue, Chicago, IL 60637}
\medskip
\centerline{(Received )}
\medskip

\centerline{\bf ABSTRACT}
\begin{quote}
Recent reanalyses of the atomic physics effects on the weak charge
in cesium have led to a value in much closer agreement with predictions of
the Standard Model.  We review precision electroweak tests, their implications
for upper bounds on the mass of the Higgs boson, possible ways in which these
bounds may be circumvented, and the requirements placed upon accuracy of
future atomic parity violation experiments by these considerations.
\end{quote}

\leftline{PACS Categories: 11.30.Er, 12.15.Ji, 12.15.Mm, 12.60.Cn}
\bigskip

The successful unification of the weak and electromagnetic interactions
\cite{GWS} has been tested to the level of radiative corrections affected by
the mass of the Higgs boson \cite{WM00}.  However, Peskin and Wells
\cite{PW01} have noted several contexts in which assumptions about electroweak
symmetry breaking can be relaxed, leading to looser bounds on the Higgs boson
mass.  As one example, a small vacuum expectation value of a Higgs triplet
\cite{FRW} can permit a Higgs boson mass in excess of 1 TeV.  Specific models
(e.g., \cite{PW01,HHT}) with this property have been constructed.  Other
related discussions may be found in \cite{DoH,CGG,MEP01}.

Among the electroweak observables that play a role in precise tests of
the radiative corrections in the theory, atomic parity violation plays a
special role.  Many types of new physics affect what are known as ``oblique
corrections,'' through vacuum polarization of the photon, $Z$, and $W$ bosons.
These effects have been described by Peskin and Takeuchi \cite{PT} in terms of
two parameters $S$ and $T$, upon which various observables depend linearly,
with $S = T = 0$ corresponding to ``no new physics,'' given nominal values of
the top quark and Higgs boson masses.  The weak charge $Q_W$ measured in
parity-violation experiments in such atoms as cesium \cite{Cs97,Cs99}, bismuth
\cite{Bi}, lead \cite{Pb}, and thallium \cite{TlS,TlO} is mainly sensitive to
the variable $S$, with very small dependence on $T$ \cite{MR,PGS,JR90}.  Thus,
atomic parity violation experiments can shed unique light on certain types of
new physics which contribute to the parameter $S$ \cite{APV95,APV97,APV99}.

Atomic physics calculations have been carried out for such systems
as cesium \cite{Csth} and thallium \cite{Tlth}.  In 1999 the JILA-Boulder
group reported measurements in cesium \cite{Cs99} that reduced uncertainties in
previous calculations.  This led to a resulting weak charge, $Q_W({\rm Cs})
= - 72.06 \pm 0.28_{\rm expt} \pm 0.34_{\rm theor} = - 72.06 \pm 0.46$ which
represented a considerable improvement with respect to previous values in this
and other atoms.  It was also more than two standard deviations away from
the Standard Model prediction \cite{MR,TLG} $Q_W({\rm Cs}) = -73.19 \pm 0.13$,
leading to speculations \cite{APV99,Cas,EL} about possible sources of the
discrepancy such as $Z'$ bosons \cite{LRR,LR} in extended gauge theories.  No
such bosons have been seen up to masses of about 600 GeV/$c^2$ \cite{CDFZp}.

Several recent contributions \cite{Der,Koz,Dzu,Sush} have re-evaluated atomic
physics corrections in cesium, paying particular attention to the Breit
interaction \cite{Br}.  Our working average for these determinations will be
$Q_W({\rm Cs}) = -72.2 \pm 0.8$.  In the present paper
we review the main electroweak
observables affecting the mass of the Higgs boson, some possible ways in which
upper bounds on this mass may be circumvented, and requirements placed upon
accuracy of future atomic parity violation experiments by these considerations.

We begin with a brief review of the formalism of \cite{PT}.  Electroweak
radiative corrections may be divided into ``oblique'' and ``direct''
contributions.  Oblique corrections (sensitive to all forms of new physics)
enter through gauge boson vacuum polarization terms, and direct corrections
include all other terms such as vertex and self-energy modifications.
At lowest order,
the $W$ mass $M_W$, the $Z$ mass $M_Z$, the electroweak couplings $g$ and $g'$,
the electric charge $e$, the weak mixing angle $\theta$, the Higgs doublet
vacuum expectation value $v$, and the Fermi
coupling constant $G_F = 1.16637(1) \times 10^{-5}$ GeV$^{-2}$ are related by
\beq
e = g \sin \theta = g' \cos \theta~~,~~~
\frac{G_F}{\s} = \frac{g^2}{8M_W^2} = \frac{g^2 + {g'}^2}{8 M_Z^2}
= \frac{1}{2 v^2}~~~
\eeq
under the assumption that the only contribution to electroweak symmetry
breaking comes from one or more Higgs doublets with vacuum expectation values
$v_i$ satisfying $\sum_i v_i^2 = v^2$.  With $\alpha \equiv e^2/4 \pi$
one then has
\beq
M_W = \frac{(\pi \alpha/\s G_F)^{1/2}}{\sin \theta}~~,~~~
M_Z = M_W/\cos \theta~~~.
\eeq
Photon vacuum polarization effects change $\alpha^{-1}$ from its value of
$\sim 137.036$ at $q^2 = 0$ to $128.933 \pm 0.021$ at $q^2 = M_Z^2$ \cite{DH}.
This important oblique correction is
sensitive to all charged particles with masses less than ${\cal O}(M_Z/2)$.

The next-most-important oblique correction arises from the large splitting
between the top and bottom quark masses \cite{MV77}, violating
a {\it custodial SU(2)} symmetry \cite{PS80} responsible for preserving
the tree-level relation $M_W = M_Z \cos \theta$.  As a result,
an effect is generated equivalent to a Higgs {\it triplet}
vacuum expectation value.
The vacuum polarization diagrams with $W^+ \to t \bar b \to W^+$ and
$Z \to (t \bar t, b \bar b) \to Z$ lead to a modification of the
relation between $G_F$, coupling constants, and $M_Z$ for neutral-current
exchanges:
\beq \label{eqn:GFr}
\frac{G_F}{\s} = \frac{g^2 + {g'}^2}{8 M_Z^2} ~~~\to~~~
\frac{G_F}{\s} \rho = \frac{g^2 + {g'}^2}{8 M_Z^2}~~~,~~
\rho \simeq 1 + \frac{3 G_F m_t^2}{8 \pi^2 \s}~~~.
\eeq
The $Z$ mass is now related to the weak mixing angle by
\beq
M_Z^2 = \frac{\pi \alpha}{\s G_F \rho \sst \cst}~~~,
\eeq
where we have omitted some small terms logarithmic in $m_t$.  A precise
measurement of $M_Z$ now specifies $\theta$ only if $m_t$ is known, so
$\theta = \theta(m_t)$ and hence $M_W^2 = \pi \alpha/(\s G_F \sst)$ is also
a function of $m_t$.

To display dependence of electroweak observables on such quantities as
the top quark and Higgs boson masses $m_t$ and $M_H$, we expand the
observables about nominal values \cite{WM00} calculated for specific $m_t$
and $M_H$.  We thereby isolate the dependence on $m_t,~M_H$, and new physics
arising from oblique corrections associated with loops in the $W$ and $Z$
propagators.
For $m_t = 174.3$ GeV, $M_H = 100$ GeV, the measured value of $M_Z$ leads to a
nominal expected value of $\sin^2 \theta_{\rm eff} = 0.23140$.  In what follows
we shall interpret the effective value of $\sin^2 \theta$ as that measured via
leptonic vector and axial-vector couplings: $\sef \equiv
(1/4)(1 - [g_V^{\ell}/g_A^{\ell}])$.

Defining the parameter $T$ by $\Delta \rho \equiv \alpha T$, we find
\beq \label{eqn:Teq}
T \simeq \frac{3}{16 \pi \sin^2 \theta} \left[ \frac{m_t^2 - (174.3
~{\rm GeV})^2}{M_W^2} \right] - \frac{3}{8 \pi \cos^2 \theta}
\ln \frac{M_H}{100~{\rm GeV}} ~~~.
\eeq
The weak mixing angle $\theta$, the $W$ mass, and other electroweak observables
now depend on $m_t$ and $M_H$.

The weak charge-changing and neutral-current interactions are probed under a
number of different conditions, corresponding to different values of momentum
transfer.  In order to account for such effects we may replace the
lowest-order relations between $G_F$, couplings, and masses by
\beq
\frac{G_F}{\sqrt{2}} = \frac{g^2}{8 M_W^2} \left( 1 + \frac{\alpha S_W}{4
\sin^2 \theta} \right)~~~,~~~
\frac{G_F \rho}{\sqrt{2}} = \frac{g^2 + {g'}^2}{8M_Z^2} \left( 1 + \frac{\alpha
S_Z}{4 \sin^2 \theta \cos^2 \theta} \right)~~~,
\eeq
where $S_W$ and $S_Z$ are coefficients representing variation with momentum
transfer. Together with $T$, they express a wide variety of electroweak
observables in terms of quantities sensitive to new physics.  (The presence of
such corrections was noted in \cite{MV77}.)  The variable $U$
defined in \cite{PT} is equal to $S_W - S_Z$, while $S \equiv S_Z$.

Expressing the new physics effects in terms of deviations from nominal
values of top quark and Higgs boson masses, we have the expression
(\ref{eqn:Teq}) for $T$, while contributions of Higgs bosons and of possible
doublets of new degenerate fermions $U$ and $D$ to $S_W$
and $S_Z$, in a leading-logarithm approximation, are \cite{KL90}
\beq \label{eqn:sz}
S_W = S_Z = \frac{1}{6 \pi} \left [ \ln \frac{M_H}{100~\g/c^2} + \sum N_c
 \right ] ~~~,
\eeq
where $N_c$ is the number of colors of the new fermions, and the sum is taken
over all such doublets.  (See \cite{KL90} for the case $m_U \ne m_D$.)

A degenerate heavy fermion doublet
with $N_c$ colors thus contributes $\Delta S_Z = \Delta S_W = N_c/6 \pi$.
For example, in a minimal dynamical symmetry-breaking (``technicolor'')
scheme, with a single doublet of $N_c = 4$ fermions, one will have $\Delta S =
2/3 \pi \simeq 0.2$.  This will turn out to be marginally acceptable under
the condition that a small impurity of Higgs-triplet symmetry breaking
is admitted, while many non-minimal schemes, with larger numbers of doublets,
will be ruled out.

The prediction $M_Z = M_W/\cos \theta$ is specific to the assumption that only
Higgs doublets of \SUL~ exist.  [SU(2)$_L$ singlets which are neutral also
have $Y=0$, and do not affect $W$ and $Z$ masses.]  For a complex $Y=2$ triplet
of the form
\beq
\Phi \equiv \left[ \begin{array}{c} \Phi^{++} \\ \Phi^+ \\ \Phi^0
\end{array} \right]~~,~~~  I_{3L} = \left\{ \begin{array}{c} +1 \\ 0 \\ -1
\end{array} \right.~~~,
\eeq
the contribution of $\vev{\Phi^0} = V_{1,-1}/\s$ to gauge boson masses
(see, e.g., \cite{StA}) is
\beq
M_W^2 = \frac{g^2}{4}(v^2 + 2 V^2_{1,-1})~~,~~~
M_Z^2 = \left( \frac{g^2 + {g'}^2}{4} \right) (v^2 + 4 V^2_{1,-1})~~~,
\eeq
so the ratio $\rho = (M_W/M_Z \cos \theta)^2$ is no longer 1, but becomes
\beq
\rho = \frac{v^2 + 2 V^2_{1,-1}}{v^2 + 4 V^2_{1,-1}}~~~.
\eeq
This type of Higgs boson thus leads to $\rho < 1$.

In the $Y=0$ triplet
\beq
\Phi \equiv \left[ \begin{array}{c} \Phi^+ \\ \Phi^0 \\
\Phi^- \end{array} \right]~~~,~~
 I_{3L} = \left\{ \begin{array}{c} +1 \\ 0 \\ -1
\end{array} \right.~~~,
\eeq
if $\vev{\Phi^0} = V_{1,0}/\s$, we find by a similar calculation that
\beq
M_W^2 = \frac{g^2}{4}(v^2 + 4 V^2_{1,0})~~,~~~
M_Z^2 = \left( \frac{g^2 + {g'}^2}{4} \right) v^2~~~.
\eeq
Here we predict
\beq
\rho = 1 + \frac{4 V_{1,0}^2}{v^2}~~~,
\eeq
so this type of Higgs boson leads to $\rho > 1$.

We now present a simplified analysis of present electroweak data in the $S$,
$T$ framework which captures the essential elements.  (See, e.g., \cite{MS01}
for a more complete version.) We shall assume $S_W = S_Z = S$.
The present analysis is an update of \cite{APV99}, which may be consulted for
further references.  (See also \cite{PW01}.)  We include
atomic parity violation in cesium and thallium (as in \cite{APV99}),
the observed values of $M_W$ as measured at the Fermilab Tevatron and LEP-II,
the leptonic width of the $Z$, the value of $\sin^2 \theta_{\rm eff}$ as
measured in various asymmetry experiments at the $Z$ pole in $e^+ e^-$
collisions, and the recent measurement by the NuTeV Collaboration \cite{NuTeV}
of a combination of neutrino and antineutrino neutral-current
to charged-current cross section ratios $R_\nu$ and $R_{\bar \nu}$.

The inputs, their nominal
values for $m_t = 174.3$ GeV and $M_H = 100$ GeV, and their dependences on $S$
and $T$ are shown in Table \ref{tab:ST}.  The value of $Q_W({\rm Cs})$
in this table has been distilled from those in Table \ref{tab:QWCs}. On the
basis of the comment in Ref.\ \cite{Sush} that other determinations have
ignored a strong-field correction, we have taken as a central value that
implied by Ref.\ \cite{Sush}.  The NuTeV data may be expressed as an effective
measurement of the $W$ mass, with small corrections quoted in Ref.\
\cite{NuTeV}.  We use these corrections to arrive at the $S$ and $T$
dependences of ``$M_W$''.  These supersede those quoted in Ref.~\cite{APV99},
which were incorrectly inferred from an earlier NuTeV report \cite{NuTeV99}.

% This is Table 1
\begin{table}[t]
\begin{center}
\caption{Electroweak observables described in fit. \label{tab:ST}}
\medskip
\begin{tabular}{c c c} \hline
Quantity      &   Experimental   &   Theoretical \\
              &      value       &    value      \\ \hline
$Q_W({\rm Cs})$ & $-72.2 \pm 0.8^{~a)}$ & $-73.19^{~b)} - 0.800 S - 0.007 T$ \\
$Q_W({\rm Tl})$ & $-115.0 \pm 4.5^{~c)}$ & $-116.8^{~d)} - 1.17S - 0.06T$ \\
$M_W~(\g/c^2)$ & $80.451 \pm 0.033^{~e)}$  & $80.385^{~f)} -0.29S + 0.45T$ \\
$\Gamma_{\ell\ell}(Z)$ (MeV) & $83.991 \pm 0.087^{~g)}$ & $84.011^{~f)} -0.18S
+ 0.78T$ \\
$\sef$ & $0.23152 \pm 0.00017^{~g)}$ & $0.23140^{~f)}
 + 0.00362 - 0.00258T$ \\
``$M_W$'' (GeV/$c^2$) & $80.136 \pm 0.084^{~h)}$ & 
 $80.385^{~f)} -0.27S + 0.56T$ \\ \hline
\end{tabular}
\leftline{$^{a)}$ Weak charge in cesium \cite{Cs97,Cs99} incorporating
recalculated atomic physics}
\leftline{\qquad corrections \cite{Der,Koz,Dzu,Sush}.}
\leftline{$^{b)}$ Calculation \cite{MR} incorporating electroweak corrections,
updated in \cite{TLG}.}
\leftline{$^{c)}$ Weak charge in thallium \cite{TlS,TlO} incorporating
atomic physics corrections \cite{Tlth}.}
\leftline{$^{d)}$ Calculation incorporating electroweak corrections \cite{PSBL}
.}
\leftline{$^{e)}$ Ref.\ \cite{DC01} $^{f)}$ Ref.\ \cite{WM00}. $^{g)}$ Ref.\
\cite{LEW01}.}
\leftline{$^{h)}$ Based on NuTeV measurement of ratios $R_\nu$ and $R_{\bar
 \nu}$ (see text) \cite{NuTeV}.}
\end{center}
\end{table}

\begin{figure}
\centerline{\epsfysize = 4in \epsffile {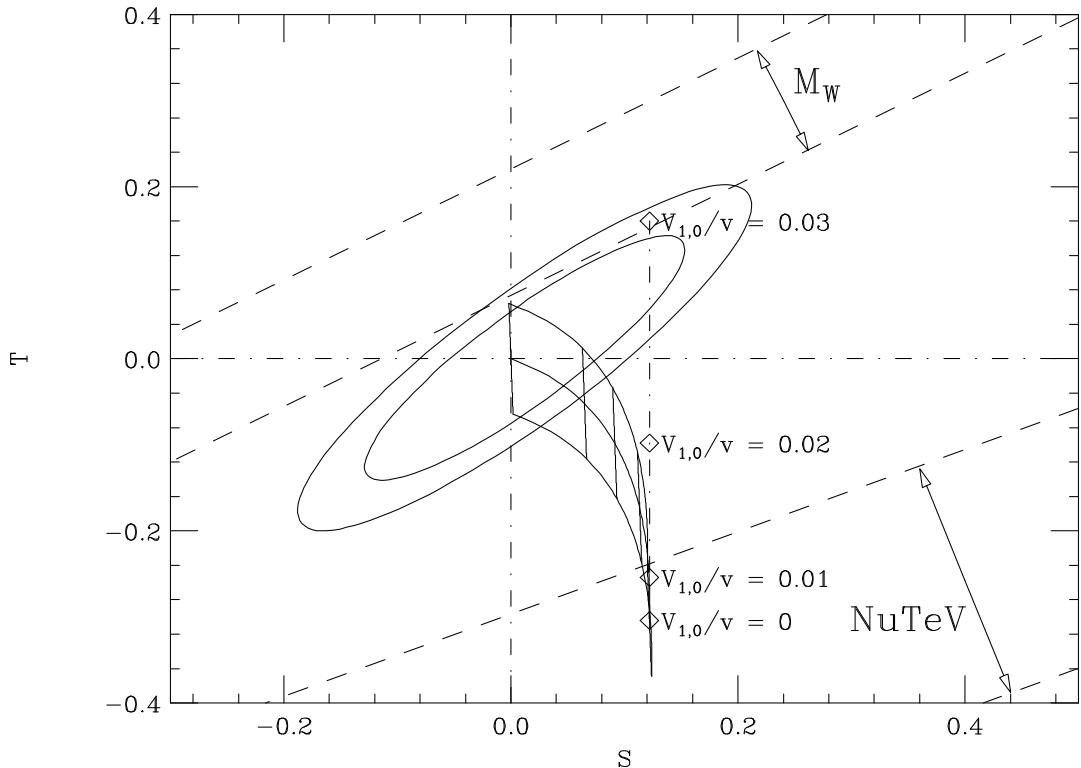}}
\caption{Regions of 68\% (inner ellipse) and 90\% (outer ellipse) confidence
level values of $S$ and $T$ based on the comparison of the theoretical and
experimental electroweak observables shown in Table \ref{tab:ST}, including
atomic parity violation data (first two lines).  Diagonal bands bounded by
dashed lines correspond to $\pm 1 \sigma$ constraints associated with
direct $M_W$ measurements (upper left) and with NuTeV measurements \cite{NuTeV}
of $R_\nu$ and $R_{\bar \nu}$ (lower right).  Standard model predictions
(solid nearly vertical lines and drooping curves) are explained in text.
\label{fig:STapv}}
\end{figure}

We do not constrain the top quark mass; we shall display its effect on
$S$ and $T$ explicitly.  Each observable specifies a pair of parallel lines
in the $S-T$ plane.  The leptonic width mainly constrains $T$; $\sef$
provides a good constraint on $S$ with some $T$-dependence; and direct
measurements of $M_W$ or values of ``$M_W$'' implied by the NuTeV data lie
in between.  The atomic parity violation experiments constrain $S$ almost
exclusively, but we shall see that they have little impact at their present
level of sensitivity.  Since the slopes are very different, the resulting
allowed region is an ellipse, shown in Figure \ref{fig:STapv} (with
the atomic parity violation data).  The corresponding figure with those data
omitted is almost identical, but shifted in central values
by $+0.01$ unit in each of $S$ and $T$.  The fits with and without the atomic
parity violation data are compared in Table \ref{tab:fits}.

% This is Table 2
\begin{table}
\begin{center}
\caption{Values of $Q_W({\rm Cs})$ used to obtain the average in Table
\ref{tab:ST}. \label{tab:QWCs}}
\begin{tabular}{c c c} \hline
Author(s) & Reference & $Q_W({\rm Cs})$ \\ \hline
Derevianko & \cite{Der} & $-72.61 \pm 0.28_{\rm expt} \pm 0.73_{\rm theor}$ \\
Kozlov \ite & \cite{Koz} & $-72.5 \pm 0.7$ \\
Dzuba \ite & \cite{Dzu} & $-72.42 \pm 0.28_{\rm expt} \pm 0.74_{\rm theor}$ \\
Milstein and Sushkov & \cite{Sush} & $\simeq -72.2$ \\ \hline
\end{tabular}
\end{center}
\end{table}

% This is Table 3
\begin{table}
\caption{Comparison of fits with and without atomic parity violation data.
\label{tab:fits}}
\begin{center}
\begin{tabular}{c c c} \hline
            & $S_0$ & $T_0$ \\ \hline
APV data    & $0.01 \pm 0.15$ & $0.00 \pm 0.15$ \\
No APV data & $0.02 \pm 0.15$ & $0.01 \pm 0.15$ \\ \hline
\end{tabular}
\end{center}
\end{table}

Figure \ref{fig:STapv} also shows predictions \cite{PW01}
of the standard electroweak theory.  Nearly vertical lines correspond, from
left to right, to Higgs boson masses $M_H = 100,$ 200, 300, 500, 1000 GeV;
drooping curves correspond, from top to bottom, to $+1 \sigma$, central, and
$-1 \sigma$ values of $m_t = 174 \pm 5.1$ GeV.

% This is Figure 1

In the standard model, the combined constraints of electroweak observables
such as those in Table \ref{tab:ST} and the top quark mass favor a very light
Higgs boson, with most analyses favoring a value of $M_H$ so low that the
Higgs boson should already have been discovered.  The standard model prediction
for $S$ and $T$ curves down quite sharply in $T$ as $M_H$ is increased,
quickly departing from the region allowed by the fit to electroweak
data.  (Useful analytic expressions for the contribution of a Higgs boson
to $S$ and $T$ are given by \cite{FRW}.)  However, if a small amount of
triplet symmetry breaking is permitted, the agreement with the electroweak
fit can be restored.  As an example, a value of $V_{1,0}/v$ slightly
smaller than 3\% permits
satisfactory agreement even for $M_H = 1$ TeV, as shown by the vertical
line in the Figure.

If electroweak-symmetry-breaking vacuum expectation values are not due to a
fundamental Higgs boson but rather to higher-dimension operators, one might
well expect both Higgs doublet and Higgs triplet contributions, with their
ratio indicating a geometric hierarchy of symmetry-breaking mass scales.  (See
\cite{LeRR,LeR} for some early examples of this behavior.)  One might then
expect Higgs singlets of various types to have characteristic vacuum
expectation values of $V_0 \simeq v^2/V_{1,0} \simeq 246~{\rm GeV}/0.03
\simeq 8$ TeV.  It is questionable whether the CERN Large Hadron Collider
(LHC), with a total $pp$ center-of-mass energy of 14 TeV, could shed light
on this mass scale.

What atomic-parity violation measurement would have a noticeable effect on the
fit shown in Figure 1?  The present error of $\pm 0.8$ on $Q_W({\rm Cs})$ is
equivalent to $\Delta S = \pm 1$.  To match the error of $\pm 0.15$ on $S$ from
the fits, one would have to determine $Q_W({\rm Cs})$ a factor of between 6 and
7 more precisely than at present.  The most significant ($>3 \sigma$)
discrepancies in present electroweak fits are (a) the difference between
values of $\sef$ measured using asymmetries of quarks and those using
leptons \cite{LEW01}, and (b) the the difference between directly measured
$M_W$ values and those inferred from the neutral-current data of NuTeV
\cite{NuTeV}.  Reduction of theoretical uncertainties associated with atomic
physics calculations will be needed before one can claim similar discrepancies
in atomic parity violation.

The need for determining $S$ independently of $T$
is highlighted by the Higgs-triplet example we have quoted.  If a small
Higgs-triplet contribution is present, one should be prepared to
determine $S$ to an accuracy of better than $\pm 0.1$ if one wishes to
pinpoint the Higgs boson mass via this indirect method.  Of course, there is
no substitute for direct searches, which the Fermilab Tevatron and the CERN
Large Hadron Collider will provide in due course.  It is also seen from
Figure 1 that a minimal ``technicolor'' contribution of $\Delta S
= 0.2$ cannot be excluded at the 90\% confidence-level limit if one is
prepared to admit a Higgs-triplet contribution and a very heavy Higgs boson.

We thank Z. Luo for discussions regarding the $S$ and $T$ dependence of the
NuTeV measurement, K. S. McFarland, O. P. Sushkov, and G. P. Zeller for helpful
comments, and Michael Peskin for communicating the curves of Ref.\ \cite{PW01}
quoted in Fig.\ 1.  This work was supported in part by the United States
Department of Energy under Grant No. DE FG02 90ER40560. 

% Journal and other miscellaneous abbreviations for references
\def \ajp#1#2#3{Am.~J.~Phys.~{\bf#1}, #2 (#3)}
\def \apny#1#2#3{Ann.~Phys.~(N.Y.) {\bf#1}, #2 (#3)}
\def \app#1#2#3{Acta Phys.~Polonica {\bf#1}, #2 (#3)}
\def \arnps#1#2#3{Ann.~Rev.~Nucl.~Part.~Sci.~{\bf#1}, #2 (#3)}
\def \cmts#1#2#3{Comments on Nucl.~Part.~Phys.~{\bf#1}, #2 (#3)}
\def \cn{Collaboration}
\def \cp89{{\it CP Violation,} edited by C. Jarlskog (World Scientific,
Singapore, 1989)}
\def \dpfa{{\it The Albuquerque Meeting: DPF 94} (Division of Particles and
Fields Meeting, American Physical Society, Albuquerque, NM, Aug.~2--6, 1994),
ed. by S. Seidel (World Scientific, River Edge, NJ, 1995)}
\def \dpff{{\it The Fermilab Meeting: DPF 92} (Division of Particles and Fields
Meeting, American Physical Society, Batavia, IL., Nov.~11--14, 1992), ed. by
C. H. Albright \ite~(World Scientific, Singapore, 1993)}
\def \efi{Enrico Fermi Institute Report No. EFI}
\def \epjc#1#2#3{Eur.~Phys.~J.~C~{\bf #1}, #2 (#3)}
\def \epl#1#2#3{Europhys.~Lett.~{\bf #1}, #2 (#3)}
\def \f79{{\it Proceedings of the 1979 International Symposium on Lepton and
Photon Interactions at High Energies,} Fermilab, August 23-29, 1979, ed. by
T. B. W. Kirk and H. D. I. Abarbanel (Fermi National Accelerator Laboratory,
Batavia, IL, 1979}
\def \hb87{{\it Proceeding of the 1987 International Symposium on Lepton and
Photon Interactions at High Energies,} Hamburg, 1987, ed. by W. Bartel
and R. R\"uckl (Nucl.~Phys.~B, Proc.~Suppl., vol. 3) (North-Holland,
Amsterdam, 1988)}
\def \ib{{\it ibid.}~}
\def \ibj#1#2#3{~{\bf#1}, #2 (#3)}
\def \ichep72{{\it Proceedings of the XVI International Conference on High
Energy Physics}, Chicago and Batavia, Illinois, Sept. 6 -- 13, 1972,
edited by J. D. Jackson, A. Roberts, and R. Donaldson (Fermilab, Batavia,
IL, 1972)}
\def \ijmpa#1#2#3{Int.~J.~Mod.~Phys.~A {\bf#1}, #2 (#3)}
\def \ite{{\it et al.}}
\def \jpb#1#2#3{J.~Phys.~B {\bf#1}, #2 (#3)}
\def \lkl87{{\it Selected Topics in Electroweak Interactions} (Proceedings of
the Second Lake Louise Institute on New Frontiers in Particle Physics, 15 --
21 February, 1987), edited by J. M. Cameron \ite~(World Scientific, Singapore,
1987)}
\def \ky85{{\it Proceedings of the International Symposium on Lepton and
Photon Interactions at High Energy,} Kyoto, Aug.~19-24, 1985, edited by M.
Konuma and K. Takahashi (Kyoto Univ., Kyoto, 1985)}
\def \mpla#1#2#3{Mod.~Phys.~Lett.~A {\bf#1}, #2 (#3)}
\def \nc#1#2#3{Nuovo Cim.~{\bf#1}, #2 (#3)}
\def \np#1#2#3{Nucl.~Phys.~{\bf#1}, #2 (#3)}
\def \pisma#1#2#3#4{Pis'ma Zh.~Eksp.~Teor.~Fiz.~{\bf#1}, #2 (#3) [JETP Lett.
{\bf#1}, #4 (#3)]}
\def \pl#1#2#3{Phys.~Lett.~{\bf#1}, #2 (#3)}
\def \pla#1#2#3{Phys.~Lett.~A {\bf#1}, #2 (#3)}
\def \plb#1#2#3{Phys.~Lett.~B {\bf#1}, #2 (#3)}
\def \pr#1#2#3{Phys.~Rev.~{\bf#1}, #2 (#3)}
\def \pra#1#2#3{Phys.~Rev.~A {\bf#1}, #2 (#3)}
\def \prc#1#2#3{Phys.~Rev.~C {\bf#1}, #2 (#3)}
\def \prd#1#2#3{Phys.~Rev.~D {\bf#1}, #2 (#3)}
\def \prl#1#2#3{Phys.~Rev.~Lett.~{\bf#1}, #2 (#3)}
\def \prp#1#2#3{Phys.~Rep.~{\bf#1}, #2 (#3)}
\def \ptp#1#2#3{Prog.~Theor.~Phys.~{\bf#1}, #2 (#3)}
\def \ptps#1#2#3{Prog.~Theor.~Phys.~Suppl.~{\bf#1}, #2 (#3)}
\def \rmp#1#2#3{Rev.~Mod.~Phys.~{\bf#1}, #2 (#3)}
\def \sci#1#2#3{Science {\bf#1}, #2 (#3)}
\def \si90{25th International Conference on High Energy Physics, Singapore,
Aug. 2-8, 1990}
\def \slc87{{\it Proceedings of the Salt Lake City Meeting} (Division of
Particles and Fields, American Physical Society, Salt Lake City, Utah, 1987),
ed. by C. DeTar and J. S. Ball (World Scientific, Singapore, 1987)}
\def \slac89{{\it Proceedings of the XIVth International Symposium on
Lepton and Photon Interactions,} Stanford, California, 1989, edited by M.
Riordan (World Scientific, Singapore, 1990)}
\def \smass82{{\it Proceedings of the 1982 DPF Summer Study on Elementary
Particle Physics and Future Facilities}, Snowmass, Colorado, edited by R.
Donaldson, R. Gustafson, and F. Paige (World Scientific, Singapore, 1982)}
\def \smass90{{\it Research Directions for the Decade} (Proceedings of the
1990 Summer Study on High Energy Physics, June 25--July 13, Snowmass, Colorado),
edited by E. L. Berger (World Scientific, Singapore, 1992)}
\def \tasi90{{\it Testing the Standard Model} (Proceedings of the 1990
Theoretical Advanced Study Institute in Elementary Particle Physics, Boulder,
Colorado, 3--27 June, 1990), edited by M. Cveti\v{c} and P. Langacker
(World Scientific, Singapore, 1991)}
\def \yaf#1#2#3#4{Yad.~Fiz.~{\bf#1}, #2 (#3) [Sov. J. Nucl. Phys. {\bf #1},
#4 (#3)]}
\def \zhetf#1#2#3#4#5#6{Zh.~Eksp.~Teor.~Fiz.~{\bf #1}, #2 (#3) [Sov. Phys. -
JETP {\bf #4}, #5 (#6)]}
\def \zpc#1#2#3{Zeit.~Phys.~C {\bf#1}, #2 (#3)}
\def \zpd#1#2#3{Zeit.~Phys.~D {\bf#1}, #2 (#3)}


\newpage
\begin{thebibliography}{99}

\bibitem{GWS} S. L. Glashow, \np{22}{579}{1961}; S. Weinberg,
\prl{19}{1264}{1967}; A. Salam, in {\it Proceedings of the Eighth Nobel
Symposium}, edited by N. Svartholm (Almqvist and Wiksell, Stockholm; Wiley, New
York, 1978), p.\ 367. 

\bibitem{WM00} W. J. Marciano, Brookhaven National Laboratory Report No.\
BNL-HET-00/04, hep-ph/0003181, to be published in Proceedings of MuMu99 --
5th International Conference on Physics Potential and Development of
$\mu^\pm \mu^-$ Colliders, San Francisco, CA, Dec.\ 1999.

\bibitem{PW01} M. E. Peskin and J. D. Wells, Stanford Linear Accelerator
Center Report No.\ SLAC-PUB-8763, hep-ph/0101342, submitted to Phys.\ Rev.\ D.

\bibitem{FRW} J. R. Forshaw, D. A. Ross, and B. E. White,
University of Manchester report MC-TH-01/07, hep-ph/0107232 (unpublished).

\bibitem{HHT} H.-J. He, C. T. Hill, and T. M. P. Tait, Univ.\ of Texas Report
No.\ UTEXAS-HEP-01-013, hep-ph/0108041 (unpublished);
H. J. He, N. Polonsky, and S. Su, \prd{64}{053004}{2001}.

\bibitem{DoH} B. Dobrescu and C. T. Hill, \prl{81}{2634}{1998}.

\bibitem{CGG} H. Collins, A. K. Grant, and H. Georgi, \prd{61}{055002}{2000}.

\bibitem{MEP01} M. E. Peskin, ``Interpretation of Precision
Electroweak Data, or Should We Really Believe There is a Light Higgs
Boson?'', seminar at Snowmass 2001 Workshop, July, 2001 (unpublished).

\bibitem{PT} M. Peskin and T. Takeuchi, \prl{65}{964}{1990}; \prd{46}{381}
{1992}. 

\bibitem{Cs97} C. S. Wood \ite, \sci{275}{1759}{1997}.

\bibitem{Cs99} S. C. Bennett and C. E. Wieman, \prl{82}{2484}{1999}.

\bibitem{Bi} M. J. D. Macpherson, K. P. Zetie, R. B. Warrington, D. N. Stacey,
and J. P. Hoare, \prl{67}{2784}{1991}.

\bibitem{Pb} D. M. Meekhof, P. Vetter, P. K. Majumder, S. K. Lamoureaux, and
E. N. Fortson, \prl{71}{3442}{1993}.

\bibitem{TlS} P. A. Vetter, D. M. Meekhof, P. K. Majumder, S. K. Lamoreaux,
and E. N. Fortson, \prl{74}{2658}{1995}.

\bibitem{TlO} N. H. Edwards, S. J. Phipp, P. E. G. Baird, and S. Nakayama,
\prl{74}{2654}{1995}.

\bibitem{MR} W. Marciano and J. L. Rosner, \prl{65}{2963}{1990}; \ibj{68}
{898(E)}{1992}.

\bibitem{PGS} P. G. H. Sandars, \jpb{23}{L655}{1990}.

\bibitem{JR90} J. L. Rosner, \prd{42}{3107}{1990}.
 
\bibitem{APV95} J. L. Rosner, \prd{53}{2724}{1996}.

\bibitem{APV97} J. L. Rosner, \cmts{22}{205}{1998}.

\bibitem{APV99} J. L. Rosner, \prd{61}{016006}{1999}.

\bibitem{Csth} V. A. Dzuba, V. V. Flambaum, and O. P. Sushkov, \pla{141}{147}
{1989}; S. A. Blundell, W. R. Johnson, and J. Sapirstein, \prl{65}{1411}{1990};
\prd{45}{1602}{1992}.

\bibitem{Tlth} V. A. Dzuba, V. V. Flambaum, P. G. Silvestrov, and O. P.
Sushkov, \jpb{20}{3297}{1987}.

\bibitem{TLG} T. Takeuchi, W. Loinaz, and A. Grant, Virginia Tech report
VPI-IPPAP-99-03, hep-ph/9904207, presented by T. Takeuchi at {\it Hadron
Collider Physics 13}, Mumbai, India, January 14--20, 1999 (unpublished).

\bibitem{Cas} R. Casalbuoni, S. De Curtis, D. Dominici, and R. Gatto,
\plb{460}{135}{1999}.

\bibitem{EL} J. Erler and P. Langacker, \plb{456}{68}{1999}.

\bibitem{LRR} P. G. Langacker, R. Robinett, and J. L. Rosner, \prd{30}{1470}
{1984}.

\bibitem{LR} D. London and J. L. Rosner, \prd{34}{1530}{1986}.

\bibitem{CDFZp} CDF \cn, F. Abe \ite, \prl{79}{2192}{1997}.

\bibitem{Der} A. Derevianko, \prl{85}{1618}{2000}; University of Nevada at
Reno Report, physics/0108033 (unpublished).

\bibitem{Koz} M. G. Kozlov, S. G. Porsev, and I. I. Tupitsyn, \prl{86}{3260}
{2001}.

\bibitem{Dzu} V. A. Dzuba, C. Harabati, W. R. Johnson, and M. S. Safronova,
\pra{63}{044103}{2001}.

\bibitem{Sush} A. I. Milstein and O. P. Sushkov, preprint hep-ph/0109257
(unpublished).
\bibitem{Br} G. Breit, \pr{34}{553}{1929}; \ibj{36}{383}{1930};
\ibj{39}{616}{1932}.

\bibitem{DH} M. Davier M and A. H\"ocker, \plb{419}{419}{1998}; \ibj{435}{427}
{1998}.

\bibitem{MV77} M. Veltman, \np{B123}{89}{1977}; \app{B8}{475}{1977}.

\bibitem{PS80} P. Sikivie, L. Susskind, M. B. Voloshin, and V. Zakharov,
\np{B173}{189}{1980}.

\bibitem{KL90} D. C. Kennedy and P. G. Langacker, \prl{65}{2967}{1990};
\ibj{66}{395(E)}{1991}.

\bibitem{StA} J. L. Rosner, \efi 01-34, hep-ph/0108195, lectures at the 55th
Scottish Universities' Summer School in Particle Physics, St.\ Andrews,
Scotland, August 7-23, 2001.  Proceedings to be published by the Institute
of Physics (U.K.).

\bibitem{MS01} M. Swartz, lecture at Snowmass 2001 Workshop,
transparencies available at http://pha.jhu.edu/~morris/higgs.pdf.

\bibitem{NuTeV} NuTeV \cn, G. P. Zeller \ite, preprint hep-ex/0110059,
submitted to Phys.\ Rev.\ Letters.

\bibitem{NuTeV99} NuTeV \cn, G. P. Zeller \ite, preprint hep-ex/9906024,
published in Proceedings of American Physical Society (APS) Meeting of the
Division of Particles and Fields (DPF 99), Los Angeles, CA, 5--9 Jan 1999.
 
\bibitem{PSBL} P. G. H. Sandars and B. W. Lynn, \jpb{27}{1469}{1994}.

\bibitem{DC01} D. Charlton, plenary talk at International Europhysics
Conference on High Energy Physics, Budapest, Hungary, July 12--18, 2001.

\bibitem{LEW01} LEP Electroweak Working Group; see web page
http://lepewwg.web.cern.ch/LEPEWWG for periodic updates.

\bibitem{LeRR} C. N. Leung, R. W. Robinett, and J. L. Rosner,
in {\em Proceedings of the Neutrino Mass Mini-Conference}, Telemark,
WI., Sept. 23-25, 1982, edited by Vernon Barger and David Cline, American
Institute of Physics, New York, 1983, p. 202.

\bibitem{LeR} C. N. Leung and J. L. Rosner, \prd{29}{2132}{1984}.

\end{thebibliography}
\end{document}